\documentclass{svproc}
\usepackage{amsmath,bm,bbm,latexsym,soul,nicefrac,young,youngtab}
\usepackage{graphicx,caption,multirow,dcolumn}
\usepackage{acronym}
\usepackage{tikz}

\usepackage{url}

\usetikzlibrary{arrows,shapes,positioning,shadows,trees}
\tikzset{
  basic/.style  = {draw, text width=5cm, drop shadow, font=\sffamily, rectangle},
  root/.style   = {basic, rounded corners=2pt, thin, align=center,
                   fill=green!30},
  level 2/.style = {basic, rounded corners=6pt, thin,align=center, fill=green!60,
                   text width=12em},
  level 3/.style = {basic, thin, align=left, fill=pink!60, text width=10em}
}

\begin{document}
\mainmatter              
\title{Gravitational Wave Astrophysics with TianQin:\\
A brief progress review}
\titlerunning{TQ GW astrophysics}  
%
\author{Yi-Ming Hu\inst{1} }
\authorrunning{Yi-Ming Hu et al.} 
%
\tocauthor{Yi-Ming Hu}
\institute{MOE Key Laboratory of TianQin Mission, TianQin Research Center for Gravitational Physics $\&$  School of Physics and Astronomy, Frontiers Science Center for TianQin, CNSA Research Center for Gravitational Waves, Sun Yat-sen University (Zhuhai Campus), Zhuhai 519082, China
\email{huyiming@sysu.edu.cn},\\
WWW home page:
\texttt{https://yiminghu-sysu.github.io/}}

\maketitle              

\acrodef{GW}{Gravitational Wave}
\acrodef{PSD}{Power Spectral Density}
\acrodef{EMRI}{Extreme Mass Ratio Inspiral}
\acrodef{SNR}{Signal-to-Noise Ratio}
\acrodef{VB}{Verification Binary}
\acrodef{EM}{Electromagnetic}
\acrodef{SGWB}{Stochastic Gravitational Wave Background}

\begin{abstract}
As a space-borne gravitational wave observatory, TianQin can observe a large variety of gravitational wave sources.
The rich signals can be composed by different types of astronomical systems, like Galactic compact binaries, inspiral of stellar mass black holes, merger of massive black holes, and extreme mass ratio inspirals.
The incoherent summation of these signals can also form a stochastic gravitational wave background. 
Using the future TianQin observation, it is possible to put stringent constraints on fundamental science, like the expansion history of our universe.
In this work, we provide a brief review of these topics that contributes to the 7th International Workshop on the TianQin Science Mission.
\keywords{Gravitational Wave, TianQin, Black Hole, White Dwarf}
\end{abstract}
\section{The TianQin Observatory}
%
TianQin is a proposed space-borne gravitational wave mission. 
Three satellites are expected to be launched in the 2030s to form the constellation.
The satellites will protect the test masses in the core, shielding them from external disturbances, so that the test masses can keep in a free-falling status and closely follow the geodesics.
High-precision laser interferometry will be used to measure the small changes in the length between test masses, which reflects the key information of the sources \cite{TianQin:2015yph}.
According to the design, the TianQin can detect gravitational waves in the milli-Hertz frequency band, covering a wide range of sources coming from a rich variety of locations, from as close as our cosmic backyard, to as far as the end of the observable universe \cite{Huang:2020,liu_hu_2020,TQ_MBH_2019,fan_hu_2020,liang_hu_2022}. 

%
The sensitivity of the TianQin can be indicated by the \ac{PSD}, which characterizes the frequency-dependant fluctuations of the noise.
For the convenience of calculation, the sky-averaged response is often absorbed in the expression of the sensitivity curve.
The effective \ac{PSD} of the TianQin $S_n(f)$ can be approximated as \cite{Huang:2020}
\begin{equation}
S_n(f) = \frac{1}{L^2} \left[ \frac{4S_a}{(2\pi f)^4}\left(1+ \frac{10^{-4}{\rm Hz}}{f} \right)+S_x \right] \times \left [ 1+ 0.6\left(\frac{f}{f_*}\right)^2 \right]
\end{equation}
Here, $L=\sqrt{3}\times 10^8 {\rm m}$ is the arm-length between any pair of satellites, $S_a = 10^{-30}{\rm m^2~s^{-4}~Hz^{-1}}$ is the acceleration noise, $S_x = 10^{-24}{\rm m^2~Hz^{-1}}$ is the displacement measurement noise, and $f_* = c/ 2\pi L = 0.28{\rm Hz}$ is the transfer frequency. 
We can observe that the \ac{PSD} can be determined by the three parameters, the acceleration noise determines the low-frequency behavior, the displacement measurement noise determines the level of the bucket, and the armlength/transfer frequency determines the high-frequency sensitivity.

%
The TianQin satellites are designed to follow a geocentric orbit, with the orbit height of $10^8{ \rm m}$.
The orbital period around the Earth is around 3.64 days, which means that the TianQin satellites will fly around 100 cycles annually.
The choice of the geocentric orbit also means that the orbital plane of the TianQin satellites is also fixed.
It is proposed that the norm of the TianQin plane is the white dwarf binary system RXJ0806.3+1527 (also known as HM Cancri, hereafter J0806).
The \ac{EM} observation of J0806 revealed its orbital frequency as well as its sky location, therefore, TianQin can use the binary system as a verification binary to calibrate the instrument performance.
J0806 has an ecliptic coordinates as $\lambda = 120.4^\circ, \beta = -4.7^\circ$.
Being close to the ecliptic planes has extra advantages, since the orbital, as well as thermal stabilities, are both close to ideal under this choice \cite{Zhang:2020paq,Ye:2024dca,Chen:2021dzg} 
However, when the direction of the Sun is too close to the orbital plane, the thermal shielding will be no longer ideal, and the detector will switch off the science mode for better protection.
This led to the ``three months on + three months off" work scheme of the TianQin mission, and the nominal duty cycle is 50\% \cite{TianQin:2015yph,TianQin:2020hid}.

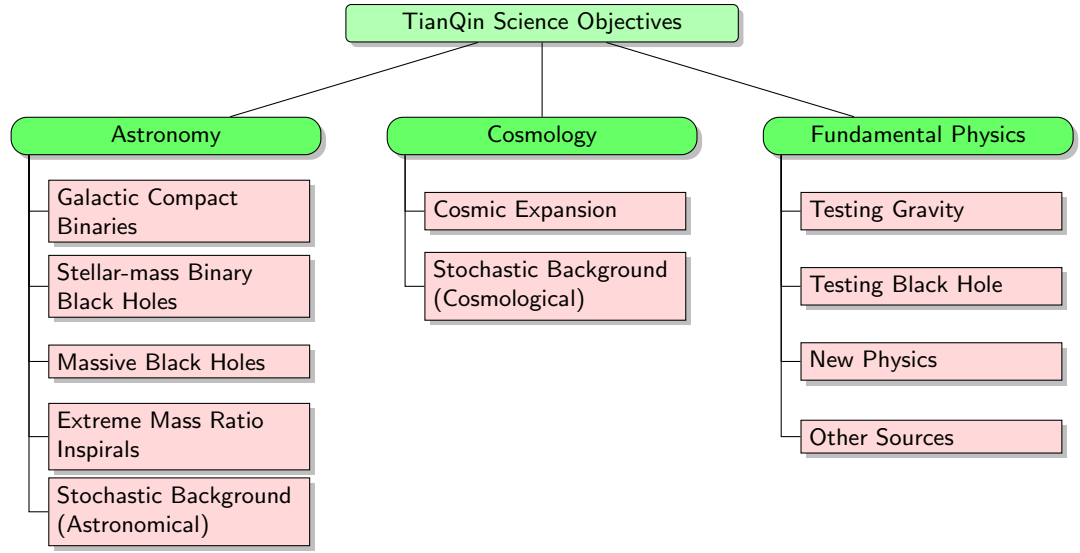
\begin{figure}[htbp] 
\begin{center}
    \begin{tikzpicture}[
  level 1/.style={sibling distance=50mm},
  edge from parent/.style={-,draw},
  >=latex]

\node[root] {TianQin Science Objectives}
  child {node[level 2] (c1) {Astronomy}}
  child {node[level 2] (c2) {Cosmology}}
  child {node[level 2] (c3) {Fundamental Physics}};

\begin{scope}[every node/.style={level 3}]
\node [below of = c1, xshift=5pt] (c11) {Galactic Compact \\ Binaries};
\node [below of = c11] (c12) {Stellar-mass Binary Black Holes};
\node [below of = c12] (c13) {Massive Black Holes};
\node [below of = c13] (c14) {Extreme Mass Ratio Inspirals};
\node [below of = c14] (c15) {Stochastic Background (Astronomical)};

\node [below of = c2, xshift=5pt] (c21) {Cosmic Expansion};
\node [below of = c21] (c22) {Stochastic Background (Cosmological)};

\node [below of = c3, xshift=5pt] (c31) {Testing Gravity};
\node [below of = c31] (c32) {Testing Black Hole};
\node [below of = c32] (c33) {New Physics};
\node [below of = c33] (c34) {Other Sources};
\end{scope}

\foreach \value in {1,...,5}
  \draw[-] (c1.188) |- (c1\value.west);

\foreach \value in {1,2}
  \draw[-] (c2.188) |- (c2\value.west);

\foreach \value in {1,...,4}
  \draw[-] (c3.188) |- (c3\value.west);
\end{tikzpicture}
\caption{Science objectives of the TianQin observatory.} 
\label{fig:SciObj}

\end{center}
\end{figure}
%
With the operation of TianQin, sources within 0.1mHz to 1Hz can be recorded by TianQin, providing a powerful tool to decipher the dark universe.
Theoretical work indicates that TianQin can observe sources near and far, light and heavy.
In the Milky Way, it is expected that there exist hundreds of millions of double white dwarfs and thousands of them can be detected by TianQin \cite{Huang:2020}.
Ground-based gravitational wave detectors have observed the late inspiral and merger of the other type of stellar-mass compact binaries, namely the stellar-mass binary black holes.
TianQin can observe the early inspirals emitted years before the merger \cite{liu_hu_2020}.
If massive black holes capture stellar-mass black holes, the heavily uneven binary will form the \ac{EMRI}, where the small black hole follows the geodesic and delivers rich information about the central black hole through complicated orbits \cite{fan_hu_2020}.
Binaries of massive black holes can emit strong \ac{GW} signals so that if the masses are ideal, even if they are on the edge of the observable universe, TianQin will be able to make reliable detections \cite{TQ_MBH_2019}.
In addition to the individual sources, the superposition of various \acp{GW} can form a noise-like stochastic background \cite{liang_hu_2022}, and their original can be either astrophysical or cosmological.
TianQin's observation of these systems will not only provide an important message about the sources, but also provide ideal laboratories to study the nature of gravity, the black hole, and the expansion of the universe. 
It might also reveal pivotal evidence for new physics beyond the standard model \cite{TianQin:2020hid}.
We summarise the major scientific objectives in Fig. \ref{fig:SciObj}.

In this document, we aim to provide a brief overview of \ac{GW} astronomy with TianQin.
We will focus on what sources can TianQin observe, the expected detection rate, and anticipated parameter estimation precision.
Each of the following sections will be dedicated to stellar-mass compact binaries, extreme mass ratio inspirals, massive black hole binaries, and stochastic gravitational wave backgrounds. 
At the end, we will briefly summarize the \ac{GW} astronomy with TianQin and the progress of the project on \ac{GW} astronomy research.

\section{Gravitational Wave Sources}
\begin{figure}[!ht]
\centering
\includegraphics[width=\textwidth]{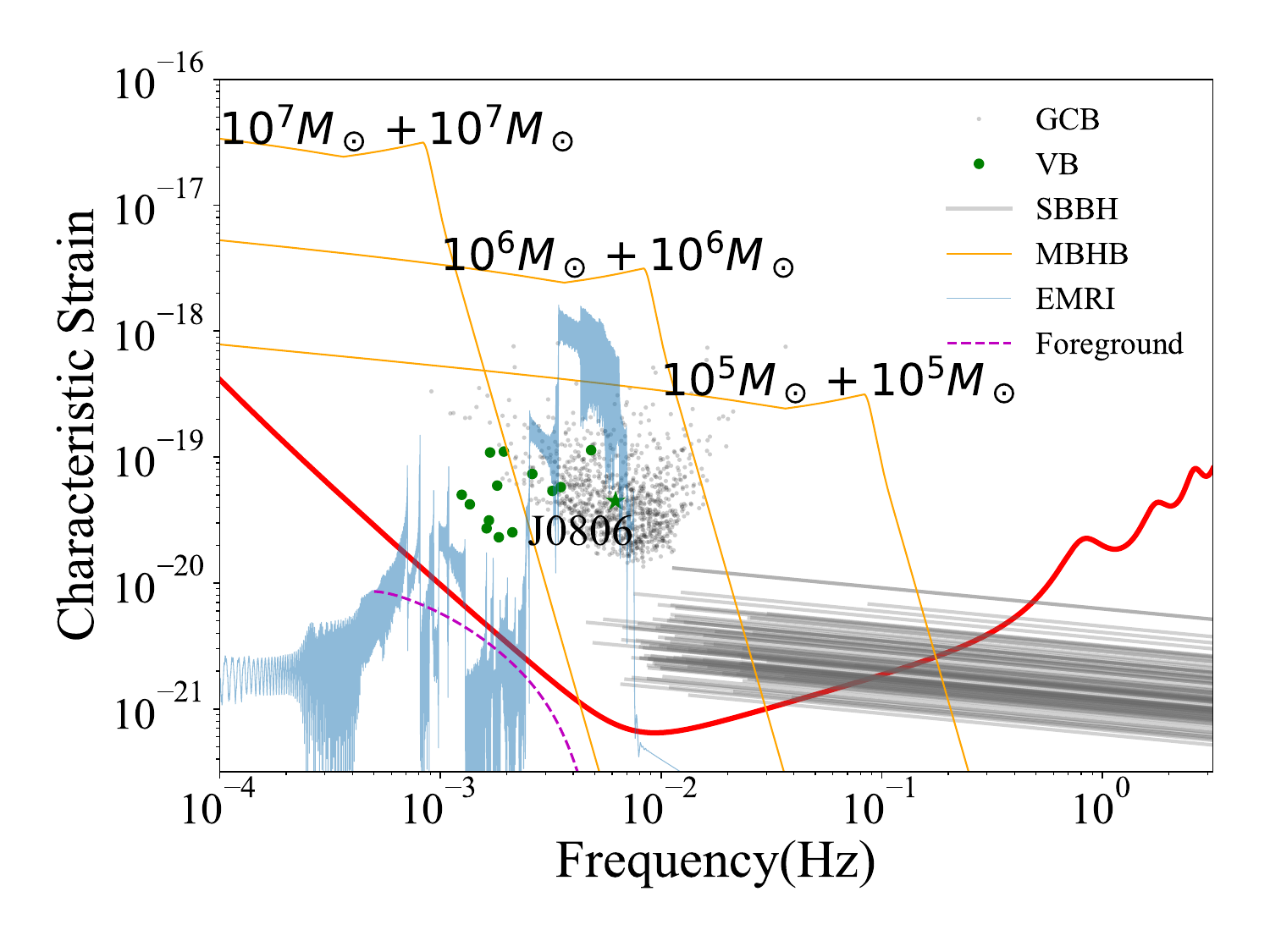}
\caption{Characteristic strain of typical sources. Gray dots represent observed white dwarf binaries through optical telescopes, while green dots indicate the strongest predicted white Galactic binaries. Gray/yellow lines indicate the inspiral (and merger) of stell stellar-mass/massive binary black holes. The blue line represents a typical extreme mass ratio inspiral, and the purple dashed line shows the foreground from the Galactic compact binaries.}
\label{fig:all_sources}
\end{figure}

\subsection{Stellar-mass Compact Binaries}
\subsubsection{Galactic Compact Binaries}

%
Observation indicates that half of the stars are in binary systems.
When the stars run out of fuel, the gravitational pull will force the stars to crash into compact objects like white dwarfs, neutron stars, or black holes, the exact form depending on the remanent mass.
Since the mass distribution of the stars is heavily tail-dominated, the vast majority of the dead stars exist in the form of white dwarfs, and those in binary systems will form double white dwarfs.
During the evolution, processes like mass transfer, common envelope phase, and/or dynamical encounter will ``harden" the binaries, namely increasing their gravitational binding energy.
Even though only a small fraction of the compact binaries in the Galaxy is expected to be so closely bound that they emit \ac{GW} signals in the mHz frequency band, the sheer volume of the total compact binaries makes the mHz \ac{GW} window packed with Galactic compact binaries. 

%
In principle, one can assess the \ac{GW} properties of the Galactic compact binaries through either theoretical modeling or astronomical observations.
However, the identification and confirmation of the binarity for Galactic compact binaries is challenging. 
Attempts have been made to make systematic searches for double white dwarfs using data from equipment like Gaia and the Zwicky transient facility.
Still, there are only dozens of \acp{VB} for TianQin, which are systems identified as double white dwarfs, and the expected \ac{SNR} larger than 5 \cite{Ren:2023}. 
The scarcity of \acp{VB} only implies the difficulty of observation through optical observation.
Using population synthesis methods, theoretical modeling of the Galactic compact binaries predicts that with a five-year mission lifetime, TianQin is expected to observe about 9,000 double white dwarfs, enriching the sample size by two orders of magnitudes and can provide precious clues for the complex processes during the late time binary evolution \cite{Huang:2020}.

%
The \ac{GW} observation from TianQin can not only contribute to observing thousands of double white dwarfs, but also provide highly precise measurement of the physical parameters.
For example, for most of the sources, the orbital periods and amplitudes can be measured to the precision of $10^{-7}$ and $20\%$, respectively.
About 40\% of the detectable sources can be localized to within 1 deg$^2$. 

Since the Galactic compact binaries can produce multi-messenger emissions, they are ideal testbeds for independent measurement of Galactic structure \cite{Korol:2018wep}, testing the theory of gravity \cite{Littenberg:2018xxx}, or even the discovery of exoplanets \cite{Tamanini:2018cqb}.

\subsubsection{Stellar-mass Binary Black Holes}
%
Since the operation of the advanced LIGO in 2015, nearly a hundred \ac{GW} events have been officially reported, with more being accumulating \cite{Chatziioannou:2024hju}.
The majority of these events are mergers of stellar-mass binary black holes.
For the ground-based detectors, these black hole mergers typically last only seconds in duration.
However, in the mHz frequency band, the evolution of the binary is much slower, so space-borne \ac{GW} observatories like TianQin can observe the early inspiral of the stellar-mass binary black holes \cite{liu_hu_2020}.
In principle, TianQin can identify the \ac{GW} signals years or even decades before the final merger, making room for well-prepared \ac{EM} follow-ups.

%
Shortly after the detection of the first \ac{GW} event GW150914, optimistic predictions were made that space-borne \ac{GW} missions can observe up to thousands of stellar-mass binary black holes \cite{Sesana:2016ljz}.
During the past decade, we have accumulated more detections to form a more comprehensive understanding of the population properties of the stellar-mass binary black holes, with careful calculations the detection rate has been updated to a lower value, that TianQin can observe about a dozen such inspirals during the five year mission lifetime \cite{liu_hu_2020}.
This means that space-borne \ac{GW} detectors can observe only nearby stellar-mass binary black holes, roughly corresponding to the redshift limit of 0.2.

%
Although the expected detection rate is not high, TianQin is still expected to make very precise measurements for the physical parameters.
For instance, even though the binaries evolve off the sensitive band long before the merger, TianQin can still predict the final merger time to the precision of the sub-second level. 
The sky location can also be pinpointed to the level of 1 square degree, which is comparable to or even smaller than many of the field of view for many survey telescopes.
The high precision in spatial and temporal parameters makes the \ac{EM} follow-up easy to implement. 
Other parameters, like the orbital eccentricity, can also be determined to the level of $10^{-4}$.
At first glance, the high precision might seem counter-intuitive, since the ground-based detectors are expected to measure the mergers with higher \acp{SNR}, yet the current detector network can only constrain the sky location to the level of tens of square degrees.
The fundamental reason lies in the fact that TianQin can accumulate signals over a long period of time.
Over the years of operation, the modulation due to Doppler shift and antenna pattern change has made TianQin very sensitive to tiny changes in the parameter space \cite{liu_hu_2020}. 

The high precision is rewarding.
For a typical source, the spatial error volume is so small that on average there is only one galaxy within, making it highly likely to identify the host galaxy for the binary black holes.
This means that even in the lack of any \ac{EM} counterpart, one can still rely on TianQin's \ac{GW} observation to draw meaningful conclusions on questions like the expansion rate of the universe \cite{Zhu:2021bpp}. 
The prospect of multi-band \ac{GW} observation also implies the possibility of constraining the environment \cite{Toubiana:2020drf},  the theory of gravity \cite{Toubiana:2020vtf}, or their formation mechanisms \cite{Liu:2021yoy}.

\subsection{Extreme Mass Ratio Inspirals}
%
Observations have revealed the ubiquitous existence of massive black holes within the center of galaxies, which are often crowded with stellar-mass compact objects.
If a stellar-mass black hole gets perturbed and becomes very close to the central massive black hole, they will form a highly unequal binary called \ac{EMRI}.
The smaller black hole follows the geodesic, and the non-linear effect predicted by the general theory of relativity makes the orbit highly complicated.
Through its long time \ac{GW} observation, it is possible to use \ac{EMRI} signals to put stringent constraints on the theory of gravity, since any tiny deviation will manifest itself over the long-term phase shift accumulation, while the \ac{GW} observation is expected to measure the phase parameters to extremely high precision. 
Therefore, \acp{EMRI} are expected to be ideal laboratories for extreme gravity.

%
The quantitative prediction on the \ac{EMRI} detection rate is quite uncertain, this is because different conditions can lead to vastly different event rates, while current observations have very weak constraints on these conditions. 
That being said, attempts have been made to quantitatively predict TianQin's detection rate.
\acp{EMRI} systems within redshift of 2 are possible to be detected with TianQin.
Studies indicate that unless in the extremely pessimistic scenario, TianQin is expected to make at least a number of \ac{EMRI} detections.
In the most optimistic case, the detection rate for TianQin can be as high as $\sim$400 per year.

%
Similar to the stellar-mass binary black holes, the TianQin can constrain the physical parameters to a very precise level.
For example, the mass parameters, the spin, and the eccentricity can all be constrained to the relative precision of $10^{-6}$.
In comparison, extrinsic parameters that only affect the amplitude are less well-constrained, for example, the sky localization ability can only be limited to 10 deg$^2$, and the luminosity distance can only be constrained to 10\%. 

The excellent precisions of intronic parameters make \ac{EMRI} systems one of the most ideal laboratories to examine gravity theories in detail.
For example, one can use TianQin's observation of \ac{EMRI} to test the no-hair theorem \cite{Zi:2021pdp,Zi:2023omh}, or to constrain the compact objects' tidal deformability \cite{Ye:2023aeo}.
The future observation of TinQin can help provide key information on \ac{EMRI} formation, which in turn could greatly improve the precision of \ac{GW} cosmology, constraining the Hubble constant to a sub-percent level \cite{Zhu:2024qpp}.

\subsection{Massive Black Hole Binaries}
%
The existence of massive black holes in the center of galaxies has been widely accepted, and the imaging of the black hole shallow has been made possible with the Event Horizon Telescope \cite{EventHorizonTelescope:2019ggy,EventHorizonTelescope:2022wkp}. 
The mergers of galaxies will bring the central massive black holes together, which eventually lead to the the merger of the binary massive black holes.
Although detailed understanding has been under debate long ago \cite{Begelman:1980vb}, the evidence that a stochastic background in the nanohertz gravitational wave window shows strong support for the ubiquitous existence of merging binary massive black holes \cite{NANOGrav:2023gor,EPTA:2023fyk,Reardon:2023gzh,Xu:2023wog}. 
In addition to the nanohertz signal, these systems can also emit strong \ac{GW} in the millihertz frequency band, especially for the merger phase as well as for the relatively lighter systems.

%
The heavier a binary system is, the stronger the \ac{GW} it emits.
For binaries of massive black holes, the strength of the \ac{GW} signal is so strong that even if a pair of massive black holes merge at the edge of the observable universe, as long as the frequency hits the sweet spot of TianQin's sensitivity band, it will still be observable with the \ac{SNR} larger than 10.
Assuming different astronomical models for the merger history as well as the growth mechanism of massive black holes, we predict that the realistic detection rate ranges from a few to a few dozen per year \cite{Wang:2019ryf}.
If TianQin collaborates with other space-borne \ac{GW} detectors like LISA, the detection rate can be further improved, especially the sky localization precision can be drastically improved \cite{Gao:2024uqc}.

%
The largest uncertainty in the rate prediction lies in the formation mechanism of massive black holes, also known as the ``seeding" mechanism. 
Traditional stellar evolution models predict that massive black holes seed from the remnant of population III stars, which is at most hundreds of solar masses.
However, this ``light seed model" encounters difficulty explaining the observed massive black holes in high redshift, which means that the black holes have to go through an extremely efficient growth stage, which could exceed the Eddington limit. 
The ``heavy seed models" are proposed to resolve this tension by introducing mechanisms like direct collapses, so that seed black holes can be born with high masses \cite{Pacucci:2023oci}. 
In order to verify and distinguish the seeding mechanism, one needs to depict the mergers at high redshift with high precision, and TianQin can do exactly that: for mergers of massive black holes happen at redshift 15, TianQin can pinpoint both the mass parameters and the distance to the level of 10\% \cite{Wang:2019ryf}.

Massive black hole mergers can also be ideal targets for performing multi-messenger observations. 
The merger of massive black holes can trigger instabilities in the accretion disks, potentially causing \ac{EM} flares.
The coordinated observation of both \ac{GW} and \ac{EM} signals can uncover critical links between the massive black holes and their host galaxies, the birth and growth of massive black holes, etc.
Notice that the \ac{SNR} accumulates slowly in the inspiral stage, and changes dramatically during the final merger stage. As \ac{SNR} increases, the localization error quickly shrinks \cite{Feng:2019wgq}.
Therefore, successful multi-messenger observation relies heavily on the ability to transmit and analyze real-time data.
Thanks to the geocentric orbit configuration, TianQin has the unique advantage of stable and low-latency data downlink, facilitated with a real-time data analysis pipeline, it is possible to coordinate observation plans with \ac{EM} telescope \cite{Chen:2023qga}.

\subsection{Stochastic Gravitational Wave Background}

%
Those \ac{GW} signals that have large enough \ac{SNR} will be resolved and analyzed individually. However, for weak signals that can not be individually resolved, their non-coherent superposition will form a stochastic background.
Even if the resolvable sources will not be matched and removed perfectly, the residual will form a foreground, especially the Galactic foreground can be comparable to the noise \ac{PSD}.
The statistical properties of the \ac{SGWB} can be indistinguishable from the instrumental noise.
Therefore, if one has no {\emph a priori} information on the instrumental noise \ac{PSD}, it is impossible to detect \ac{SGWB} from a single detector.
In the literature of \ac{SGWB} observation based on LISA, the {\emph null channel} method is proposed.
It utilized the fact that a triangle detector has three readouts, while the \ac{GW} only has two polarizations.
The other method is to use multiple independent detectors and cross-correlate their data.
The \ac{SGWB} from different detectors will have non-zero correlations, while the noise is expected to be not correlated.

%
The origin of \ac{SGWB} can be either astrophysical or cosmological.
In the millihertz band, the astrophysical \ac{SGWB} is composed of all unresolvable sources mentioned above, while cosmological \ac{SGWB} can include first-order electroweak phase transition, primordial \ac{GW}, inflation, and cosmic defects.
We limit the scope of this document to the astrophysical origin \ac{SGWB}, and leave the discussion of the cosmological \ac{SGWB} for future studies.
Since the majority of the unresolvable Galactic binaries are crowded around 1mHz, and TianQin's sensitivity is optimal for slightly higher frequencies, this means that the foreground is a less severe issue for TianQin compared with LISA.
After a five-year mission lifetime, the strongest Galactic compact binaries will be identified and removed, and the remaining will be lower than the noise \ac{PSD}.
That being said, using either the null channel or the cross-correlation method, TianQin can observe strong foreground, and the \ac{SNR} for \ac{SGWB} from stellar-mass binary black holes can reach 10.

If one relies on the null channel method, then the uncertainty of the noise \ac{PSD} has to be very small, which might not be realistic \cite{Muratore:2023gxh}.\
By joining the game, TianQin enables the possibility of cross-correlation, opening the window for reliable analysis of the \ac{SGWB}. 
TianQin will also be a game changer in terms of the \ac{SGWB} anisotropy analysis. For certain multipole components, the TianQin-LISA network will have several orders of magnitude improvement over a single detector.


\section{Summary and Discussion}

%
In summary, TianQin can observe a rich range of \ac{GW} signals, ranging from the local Galactic compact binaries, the nearby stellar-mass binary black holes, the extreme mass ratio inspirals to an intermediate distance, and the massive black hole binary mergers to the edge of the observable universe.
In addition to the individually resolvable sources, TianQin can also detect their superposition as a stochastic gravitational wave background. 
The detection rate can be highly uncertain and can differ by orders of magnitude, but it's highly likely that TianQin will observe a large number of sources, probably dominated by Galactic compact binaries.
Using these TianQin detections, one can answer some of the puzzling questions in fundamental physics and astronomies, like constraining the formation channels of the \ac{GW} sources, exploring their immediate environment, testing the general theory of relativity, probing the expansion history of the universe, etc. 

%
On the basis of the previous studies, a team of astronomers has worked on a TianQin astrophysics whitepaper for two years. It is now in a mature status, and we expect to submit the whitepaper in a short time.

%
The TianQin team does not only focus on the \ac{GW} astronomy but also pays attention to data analysis skills.
So far, a first round of studies has been performed to acquire the simulation ability \cite{Li:2023szq} and to gather basic analysis tools for different sources \cite{Gao:2024uqc,Chen:2023qga,Lu:2022ywf,Zhang:2022xuq,Cheng:2022vct,Wang:2023tle,Lyu:2023ctt,Ye:2023lok,Zhang:2024fka}
In the future, we plan to make the tool more reliable, more robust, and deal with more realistic issues \cite{Wu:2023rpn,Wang:2024bod,Liang:2024tgn}.

%
%
\bibliographystyle{unsrt}
\bibliography{ref}
\end{document}